\documentclass[letters,usenatbib]{mnras}

\usepackage{newtxtext,newtxmath}

\usepackage[T1]{fontenc}

\usepackage{natbib}
\usepackage{graphicx}	
\usepackage{amsmath}	
\usepackage{IEEEtrantools}
\usepackage{float}

\usepackage{titlecaps}

\begin{document}
\label{firstpage}


\title{An Approximation for the Capture Radius of Gaseous Protoplanets}
\author[Claudio Valletta \& Ravit Helled]{
Claudio Valletta,$^{1}$\thanks{E-mail: claudiovalletta92@gmail.com}
Ravit Helled,$^{1}$
\\
$^{1}$Center for Theoretical Astrophysics \& Cosmology, Institute for Computational Science, University of Zurich, Zurich, Switzerland\\
}

\date{Accepted 2021 July 23. Received 2021 July 19; in original form 2021 June 10
}

\maketitle


\begin{abstract}
Determining the heavy-element accretion rate of growing giant planets is crucial for understanding their formation and bulk composition. 
The solid (heavy-element) accretion rate should be carefully modeled during the various stages of giant planet formation and therefore, the planetary capture radius must be determined. 
In some simulations that model the heavy-element accretion rate, such as in N-body simulations,  the presence of the gaseous envelope is either neglected, or treated in an over-simplified manner. 
In this paper, we present an approximation for the capture radius that does not require the numerical solution of the stellar structure equations. 
Our approximation for the capture radius works extremely well for various planetesimal sizes and compositions. We show that the commonly assumed constant density assumption for inferring the capture radius leads to a large error in the calculated capture radius and we therefore suggest that our approximation should be implemented in future simulations. \\

\end{abstract}


\section{Introduction} \label{sec:introduction}
In the \textit{core accretion} model \citep{pollack96}, the standard model for giant planet formation, the growth of a giant planet begins with the formation of a heavy-element core \citep{Helled14}.
Once the core reaches about Mars' mass, its gravity is sufficiently high to accrete hydrogen and helium (hereafter, H-He) gas from the proto-planetary disk. 
The proto-planet continues growing by accreting both solids (heavy elements), in the form of planetesimals (0.1 - 100 km sized objects), and H-He until the so-called crossover mass is reached \citep{lissauer2009,Ginzburg2019} and runaway gas accretion takes place. In this last stage the planet detaches from the disk and  collapses under the influence of its gravity. 

The heavy-element accretion rate can be modeled using Safronov's equation \citep{saf2}, which is a useful approximation for the accretion of the surrounding planetesimals onto a planetary embryo:
\begin{equation}
\frac{dM_\mathrm{solid}}{dt}  =  \dot M_\mathrm{core} =
 \pi R_\mathrm{capt}^2
 \sigma_s \Omega  F_g,
\label{eq:accrete}
\end{equation} 
where $\pi R_\mathrm{capt}^2$ is the geometrical capture cross section, where $R_\mathrm{capt}$ is the capture radius, $\Omega$ is the orbital frequency, $\sigma_s$ is the solid surface density in the disk, and $F_g$ is the gravitational enhancement factor for the cross section with respect to the geometrical value, as computed by \cite{Greenzweig92}. 
If the proto-planet does not hold a gaseous envelope $R_\mathrm{capt}$ equals the core's radius. The presence of the envelope increases the capture radius by orders of magnitude \cite[e.g.,][]{pollack96}.
The solid accretion rate depends on $R_\mathrm{capt}^2$, therefore it is important to have a realistic estimate of the capture radius. 
\vspace{5cm}

A key goal of giant planet formation theory is to explain the bulk composition of giant planets. 
In order to predict the planetary composition, the total mass of heavy elements that are captured during the planetary growth must be computed accurately. 
This is, however, not an easy task because the heavy-element accretion rate depends on two distinct physical aspects: the orbit of the planetesimals disk under the effect of the gravitational force exerted by the planet \citep{Shibata1,Shibata2} and the interaction between planetesimals and the growing proto-planet \citep{Inaba03,Iaroslavitz2007,Brouwers2017,Valletta2019}. 
The former is typically computed using N-body simulations  \citep[e.g.,][]{Shibata1,podolak20}, while the planet's structure is modeled via planetary codes that solve numerically the structure equations \citep[e.g.,][]{Valletta2020,Vazan18,Helled14-2}.

N-body simulations compute accurately the orbits of planetesimals around the central star and the proto-planet. However, the envelope's structure and the capture radius are typically modeled in an over-simplistic way.
For example, \cite{Shibata1} simulated the orbits $10,000$ of planetesimals including the gravitational presence of the proto-Jupiter, assuming {\it in-situ} formation, in its late formation stage. They also included the gravitational effect of the proto-planet on the gas disk as well as gap opening. 
It was found that the mass of planetesimals captured by proto-Jupiter during the gas accretion is $\sim$ 3 M$_\oplus$. 
It was then concluded that a very high solid surface density is required to explain the total heavy-element mass in Jupiter. 
In \cite{Shibata2} the same computational method was used to estimate the amount of captured planetesimals when planetary migration is included. It was found that the amount of captured solids increases with increasing migration distances. 

However, in both of these recent studies a naive assumption for the capture radius was used. 
It was assumed that the capture radius is given by the proto-planetary mass and an a-priori assumed density, following the relation $M=4/3\pi \rho R_\mathrm{capt}^3$, where $R_\mathrm{capt}$, $\rho$ and $M$ are the capture radius, the planet's density and planet's mass, respectively. 
This is a rough approximation because while the planet is forming and gas is accreted the proto-planet density changes significantly. This can have important consequences on $R_\mathrm{capt}$.

In order to predict correctly the capture radius one should compute the trajectory of the planetesimals, solve the stellar structure equations within the proto-planet's envelope and take into account the drag force that the atmosphere exerts on the planetesimals.
\cite{Inaba03} derived an approximate expression for the capture radii of proto-planets. They showed that $R_\mathrm{capt}$ is significantly larger than the core's radius and that it mostly depends on the planetesimal's size, where $R_\mathrm{capt}$ is larger for smaller planetesimals. 
However, their expression for $R_\mathrm{capt}$  requires the knowledge of how the density change within the planet's envelope, i.e. it requires the numerical solution of the structure equation within the proto-planet. Therefore the formulation by \cite{Inaba03} can not be implemented in works similar to \cite{Shibata1}.

In this paper, we derive an approximation for $R_\mathrm{capt}$ that does not require a numerical solution of the structure equations. 
This approximation is valid throughout the entire  planetary growth: when the planet is attached to the disk, and when it is detached from the disk and runaway gas accretion takes place. 
The approximation presented here is useful for models  that do not directly compute the  structure of the planetary envelope, such as N-body simulations. 
In addition, using this approximation would significantly improve the prediction of the heavy-element accretion rate in comparison to calculations where the planet is assumed to have a constant density.

\section{Methods}
\subsection{The attached phase}
Initially, the proto-planet is attached to the disk and the outer pressure, temperature and density are determined by its orbital location. 
Following \cite{lissauer2009} we define the planet's outer radius as:
\begin{equation}
\label{radius}
R_0 = \frac{G M_P}{c_s^2+\frac{GM_P}{0.25 R_H}},
\end{equation}
where M$_P$ is the planet's mass, $c_s$ is the sound speed in the disk and R$_H$ is the Hill's radius. 
In this early phase, the gas accretion rate is obtained by demanding that the computed radius of the planet is equal to Eq.~\ref{radius}.
The structure equations of mass conservation, hydrostatic balance,
thermal gradients, and energy conservation that regulate the envelope's structure are given by:
\begin{equation}
\begin{split}
\frac{dm}{dr} =& 4 \pi r^2 \rho  \\
\frac{dP}{dr} =& -\frac{Gm}{r^2} \rho\\
\frac{dT}{dr} =& \nabla \frac{T}{P} \frac{dP}{dr}\\
\frac{dL}{dr} =& 4 \pi r^2 \rho \bigg( \epsilon - T \frac{\partial S}{\partial t}\bigg), 
\end{split}
\label{structure}
\end{equation}
where $m$ and $r$ are the mass and radius coordinates, $\rho$, $P$ and $T$ are the density, pressure, and temperature within the envelope, $L$ and $S$ are the luminosity and the entropy and the third equation defines the temperature gradient $\nabla \equiv dln t / d ln P$.
We denote the outermost pressure, temperature, and density by $P_0$, $T_0$ and $\rho_0$. The approximation of the capture radius depends on these values, which are parameters in our formula. To provide a comparison with the numerical computation we follow \cite{piso15} to compute their value.

The capture radius is determined by the gas drag exerted on the planetesimals, which is affected by the density profile within the envelope. 
Therefore, in order to estimate $R_\mathrm{capt}$ the density profile of the atmosphere needs to be determined. 
In particular, is important to have a good approximation for the density profile in the outer layers that have a larger influence on the planetesimal's orbit in comparison to the deep interior. 
We therefore focus on inferring the density profile in the outer regions of the planet.

As shown by various studies, radiation transports the heat in the outer layers of the proto-planet \citep{piso15,brouwers2020}. 
In this zone the temperature is nearly constant, while the pressure and density increase exponentially towards the planetary center. 
We can therefore represent the temperature, pressure, and density of the envelope by: 
\begin{equation}
\begin{split}
T(r) =& \; T_0  \\
P(r)=& \;P_0 \exp\bigg(\alpha(R_0/r - 1 )\bigg)\\
\rho(r) =& \; \rho_0\exp\bigg(\alpha(R_0/r - 1 )\bigg)\\.  
\end{split}
\label{ansatz}
\end{equation}
Substituting Eq.\ref{ansatz} in the second equation of Eq.\ref{structure} leads to: 
\begin{equation}
\alpha = \frac{GM \rho_0}{P_0 R_0},
\end{equation}
where we assume that the mass coordinate $m$ is equal to the total mass of the planet $M$, which is a good assumption in the outer layers of the planet.  
The density is therefore given by: 
\begin{equation}
    \rho(r) = \; \rho_0\exp\bigg(\alpha(R_0/r - 1 )\bigg).
    \label{densityapproximation}
\end{equation}

\begin{figure}
	\centering
	\includegraphics[width=0.5\linewidth]{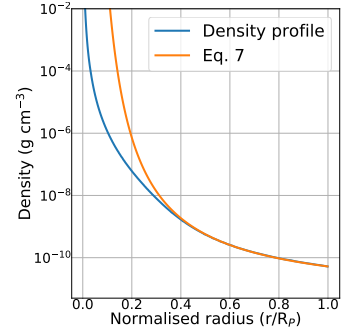}
	\caption{Density as a function of the normalised radius ($r/R_P$) within the proto-planet's envelope. The blue and orange lines correspond to the density obtained numerically and using our approximation, respectively.  
	The orbital location is assumed to be 5.2 AU and the total mass of the planet is 20 M$_\oplus$. 
	}
	\label{Fig1-density}
\end{figure}

Figure~\ref{Fig1-density} shows the comparison between  Eq.~\ref{densityapproximation} and the density obtained by solving numerically the planetary structure equations. All the numerical solutions presented in this paper are obtained using the updated model for giant planet formation presented in \cite{Valletta2020}.  
It can be seen that in a large part of the envelope the result is extremely good. 
Near the planet's center, the ansatz used to derive Eq.~\ref{densityapproximation} is no longer valid because the behavior of the density is not exponential.
The Eq.~\ref{densityapproximation} could be included in N-body simulations to include the presence of the planetary  envelope in the equation of motion of planetesimals.
In Appendix \ref{appendix1} we show further comparisons for various planetary masses and orbital locations.

We can use Eq.~\ref{densityapproximation} to derive an approximation for the capture radius.
Inserting Eq.~\ref{densityapproximation} in equation 18 of \cite{Inaba03} we find that: 


\begin{equation}
    R_\mathrm{capt} = \frac{R_0}{1+\frac{1}{\alpha} \ln \big( \frac{\rho_\star}{\rho_0}\big)}, 
    \label{finalapproximation}
\end{equation}
where 
\begin{equation}
\label{rhostar}
    \rho_\star = \frac{2 r_p \rho_p}{3 D R_H},
\end{equation}
and $r_p$ and $\rho_p$ are the planetesimal's size and density, respectively.
$D$ represents the drag coefficient defined in Eq.~11 of \cite{Inaba03}. The drag coefficient is computed according to \cite{podolak88} and for our conditions $D$ is close to unity, and therefore we use $D = 1$ in the calculation.
In principle, the value of $R_\mathrm{capt}$ also depends on the initial velocity of the planetesimals. 
In Equation~\ref{finalapproximation} it is assumed that the initial planetesimal velocity  is much smaller than the escape velocity from the protoplanet. We find that the initial velocity has almost no effect on the capture radius.   
In order to derive an approximation for $R_\mathrm{capt}$ considering different initial velocities it  is sufficient to insert Equation \ref{densityapproximation} in Equation 17 of \cite{Inaba03}. 
However, in this case there is no longer an analytic solution for R$_\mathrm{capt}$.  

\begin{figure}
	\centering
	\includegraphics[width=1\linewidth]{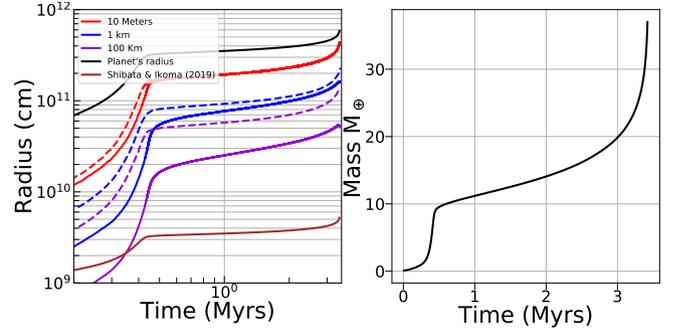}
	\caption{\textbf{Left:} R$_\mathrm{capt}$ vs.~time for various planetesimal sizes. The blue, brown, and purple curves correspond to planetesimal sizes of 10 meters, 1 km, and 100 km, respectively. The solid line indicates our Eq.~\ref{densityapproximation}, while the dashed line corresponds to $R_\mathrm{capt}$ calculated using the approximation of \protect\cite{Inaba03} and using the numerical solution for the density profile. Finally, the brown line represents the commonly used constant-density assumption \protect\cite{Shibata1}. \textbf{Right:} Planetary mass vs.~time assuming the planet is growing by accreting 100-km-sized planetesimals.
	}
	\label{Fig2-5.2au}
\end{figure}

In Appendix \ref{composition} we investigate the sensitivity of our results to the assumed planetesimal's composition.
Figure~\ref{Fig2-5.2au} shows a comparison between Eq.~\ref{finalapproximation} (solid-line) and $R_\mathrm{capt}$ computed numerically with the formula of \cite{Inaba03} (dashed-line).
We find that for small planetesimals our approximation is excellent, while the agreement is less good when assuming large planetesimals. However, the difference is typically within a factor of two. The brown line indicates the constant density approximation used in \cite{Shibata1}. There are orders of magnitude of difference between this assumption and the numerical result. 
In the right panel of Fig.~\ref{Fig2-5.2au} we show the planet's mass as a function of time, when accreting 100-km-sized planetesimals.
We use the heavy-element accretion rate of \cite{pollack96}. 
The solid accretion rates depends on $R_\mathrm{capt}^2$. It is clear from the figure that using the constant density approximation leads to unrealistic estimates of the capture radius (by orders of magnitude), which in turn affects the planetary growth and therefore should not be applied. 


\subsection{The detached phase}

\begin{figure}
	\centering
	\includegraphics[width=0.9\columnwidth]{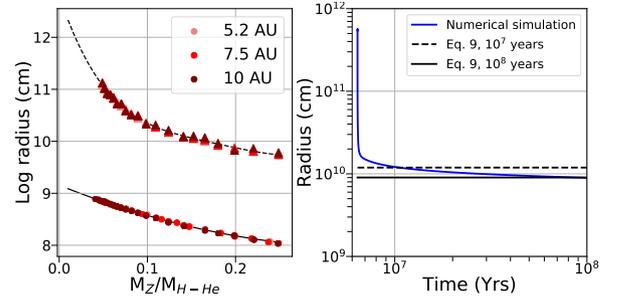}
	\caption{\textbf{Left:} Planet's radius as a function of M$_Z$/M$_{H-He}$. Each triangle and dot represent the radius of the planet after 10$^7$ and 10$^8$ years from its formation, respectively. Different colours represent different orbital's locations. \textbf{Right:} The planet's radius computed numerically (blue-line) and Eq.~\ref{detachedapproximation} (black-line) as a function of time.
	}
	\label{Fig3-runawayphase}
\end{figure}

When the gas accretion rate is limited by the maximum that the disk can supply \citep{lissauer2009} the proto-planet detaches from the disk, and as a result, the approximation derived in Equation~\ref{finalapproximation} is no longer appropriate. 
We assume that the detached phase begins at the crossover mass, i.e. when the total mass of heavy-element equals the total mass of H-He \citep{brouwers2020}.
In this phase the radius collapses very rapidly and then slowly decreases over time. After the planet reaches the point at which M$_Z$ = M$_{H-He}$, $R_\mathrm{capt}$ can be approximated as constant. 
We show in Appendix~\ref{appendix2} that the radius does not depends on the the runaway gas accretion rate. Instead, it depends  on the ratio between the total mass of heavy-elements and the mass of H-He.  
The left panel of Fig.~\ref{Fig3-runawayphase} shows the planet's radius for different values of M$_Z$/M$_{H-He}$. 
Each triangle and dot  indicate the value of the radius after 10$^7$ and 10$^8$ years from the beginning of the simulation, respectively.
The different colors represent different orbital's locations. The black lines correspond to our fit.  
The planet's radius decreases with increasing M$_Z$/M$_{H-He}$ and decreases with time.
We find that there is no difference between the various colours, and we therefore conclude that the value of the radius is independent of the assumed orbital's location.  
It should be noted, however, that we have neglected the radiation from the star. This can have important effects for close-in giant planets ($< 0.5$AU) and we hope to include this effect in future research.

The fit corresponding to Figure \ref{Fig3-runawayphase} is given by
\begin{equation}
\label{detachedapproximation}
    R = \bigg(\sum_{i=0}^{4} R_i \frac{M_Z^i}{M_{H-He}^i}\bigg)\times 10^9 cm, 
\end{equation}
where $R_0 = 12.80662188$ , $9.15426162$,  $R_1 = -50.86303789$ , $-6.74548399$, $R_2 = 382.66267044$ , $9.40271959$, $R_3 = -1388.57741163$ , $0$ and $R_4 = 1902.60362959$ , $0$. The former and latter values of the coefficients correspond to the 10$^7$ years and to the 10$^8$ years fit, respectively.
Between 10$^7$ and 10$^8$ years the radius can be can be derived with a logarithmic interpolation between these two values.
The right panel of Fig.~\ref{Fig3-runawayphase} shows the planet's radius vs.~time during the detached phase, comparing Eq.~\ref{detachedapproximation} with the radius computed numerically.
It can be seen that after a short time at which the planet shrinks rapidly, the radius can be approximated as a constant. 
Eq.~\ref{detachedapproximation} is an approximation. The planet's radius depends on several other quantities such as the total planetary mass and the core's luminosity. 
However, the dominating property is  the ratio between the heavy-element and H-He mass, M$_Z$/M$_{H-He}$  \citep[e.g.,][]{Fortney07, Baraffe08}.

\section*{Results}

In this section we present the results for the entire planetary growth.  
We suggest that $R_\mathrm{capt}$ should be approximated as follows:

\begin{equation}
R_\mathrm{capt} = 
\begin{cases}
\frac{R_0}{1+\frac{1}{\alpha} \ln \big( \frac{\rho_\star}{\rho_0}\big)}  & if M_Z > M_{H-He}, \\
\text{Eq.}\:9  \: & \text{otherwise}
\end{cases}
\label{captureradius}
\end{equation}

\begin{figure}
	\centering
	\includegraphics[width=1\linewidth]{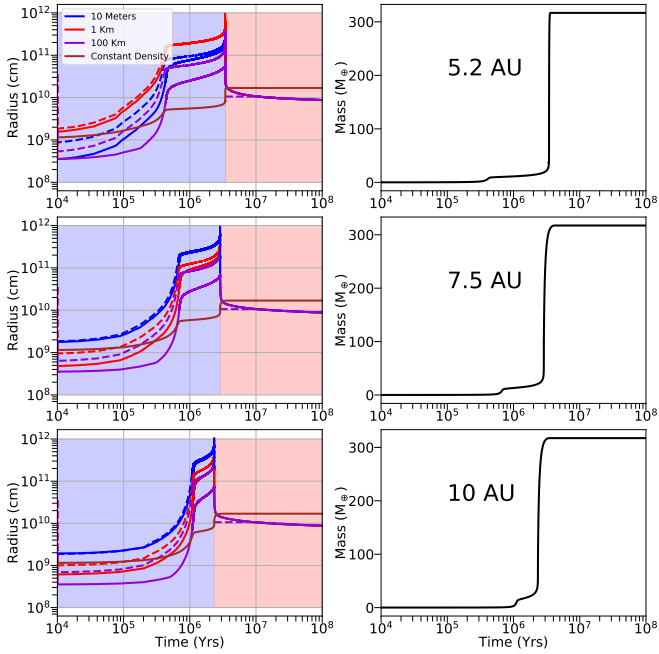}
	\caption{\textbf{Left:} R$_\mathrm{capt}$ as a function of time for different planetesimal sizes. The blue, red and purple curves correspond to sizes of 10 meters, 1 km and 100 km, respectively. The brown curve represents the constant density approximation used in \protect\cite{Shibata1}. The solid line shows $R_\mathrm{capt}$  computed with Eq.~18 of \protect\cite{Inaba03} using the density profile obtained numerically, while the dashed line represents our Eq.~\ref{captureradius}.
	The blue and red shaded regions indicate the attached and detached phase, respectively. \textbf{Right:} Planetary mass as a function of time. The various lines correspond to different formation locations.
	}
	\label{Fig4-mainplot}
\end{figure}

Figure \ref{Fig4-mainplot} presents the comparison between Eq.~\ref{captureradius} and $R_\mathrm{capt}$ computed with Eq.~18 of \cite{Inaba03} using the density computed numerically with our code \citep{Valletta2020}. 
Different dots' colours in the plot indicate different formation  locations of 5.2 AU, 7.5 AU, and 10 AU. 
The right panel plot shows the planet's mass vs.~time.  
We find that during most of the formation timescale there is a very  good agreement between Eq.~\ref{captureradius} and the numerical results. 
In particular, during the detached phase there is a small difference between the two results. 
When the planet is attached to the disk the approximation is extremely accurate for small and intermediate-size planetesimals (size $\leq$ 1 km), while for large planetesimals there is a difference of $\sim$ a factor of two. 
However, it can be seen that for all the planetesimal sizes the result is more accurate than the constant density approximation which is often used in formation models. 

\section{Summary \& Conclusions}

In order to have robust theoretical predictions for the bulk composition of giant planets, the heavy-element mass accreted during the planetary formation should be estimated. 
A key parameter in determining the heavy-element mass in growing giant planets is the capture radius. 
The presence of the gas increases significantly the capture radius and as N-body simulations typically do not account for the planetary structure it is important to find an appropriate approximation for $R_\mathrm{capt}$ accounting for the gaseous envelope. 

In this study we present an approximation for $R_\mathrm{capt}$ which is valid throughout the different stages of the planetary growth and can easily be implemented in N-body simulations. This approximation does not require the solution of the stellar structure equations.
It can be used to track consistently the formation of a giant planet in an N-body code, modeling the change in the capture radius as more and more gas is accreted. This can not be done when assuming that the planet's density remains constant because the accretion of gas decreases significantly the planet density by increasing its radius.
We divide the planetary formation into two phases: (1) when the planet is attached to the disk (M$_Z$ $>$ M$_{H-He}$), and (2) when the planet detaches from the disk (M$_{H-He}$ $>$ M$_{Z}$). 
We compare our approximation with numerical results and find that the approximation is working extremely well and suggest that this approximation should be implemented in giant planet formation models in order to correctly infer the heavy-element accretion rate of growing giant planets. 

\section*{Acknowledgements}
 The authors acknowledge  support from SNSF grant \texttt{\detokenize{200020_188460}} and the National Centre for Competence in Research ‘PlanetS’ supported by SNSF.
 
 \section*{DATA AVAILABILITY}
 The data underlying this article will be shared on reasonable request to the corresponding author.
 
  \vspace{-0.2 cm}

\bibliographystyle{mnras}
\bibliography{bibliography.bib}

\appendix
\vspace{-0.5cm}
\section{An approximation for the Density Profile of the Envelope}
\label{appendix1}
\begin{figure}
	\centering
	\includegraphics[width=0.88\linewidth]{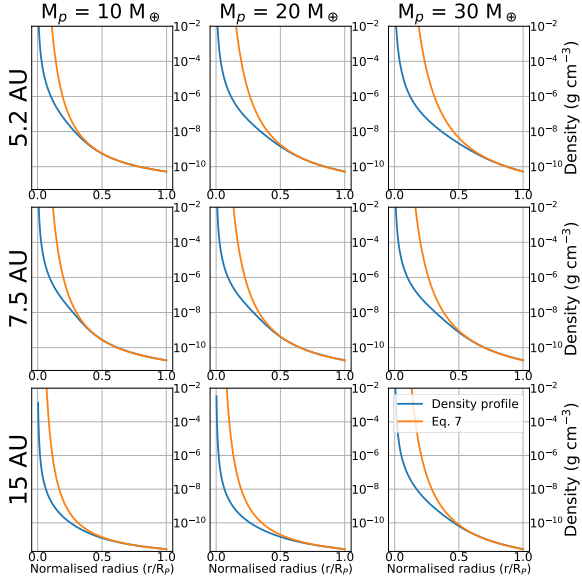}
	\caption{Density as a function of the normalised radius $r/R_P$. The blue line represents the numerical result, while the orange line corresponds to Eq.~\ref{densityapproximation}. Different lines in the figure represent different orbital locations, while different columns indicate different total planetary masses.}
	\label{Fig5-density}
\end{figure}

In this section we present further comparisons between Eq.~\ref{densityapproximation} and the density profile computed numerically.
In Fig~\ref{Fig5-density} the blue line indicates the density profile computed numerically, while the orange line represents Eq.~\ref{densityapproximation}. Different lines represent different orbital's locations, while different columns represent different total planetary masses.  
In all the cases there is an excellent agreement between the two results in the outer $\sim$ half of the planet, while for normalised radius less than 0.5 the two lines are very different. In this region the ansatz used to derive Eq.~\ref{densityapproximation} is not valid. 

\section{Dependence of $R_\mathrm{{\lowercase{capt}}}$  on the Planetesimal's composition}
\label{composition}

All the results presented above correspond to icy planetesimals with a density of 1 g cm$^{-3}$. Clearly, it is also important to investigate how $R_\mathrm{capt}$ changes with the assumed planetesimal's composition. The planetesimal's density affects the capture radius via the drag force exerted by the envelope's gas on the planetesimal.  

\begin{figure}
	\centering
	\includegraphics[width=0.85\linewidth]{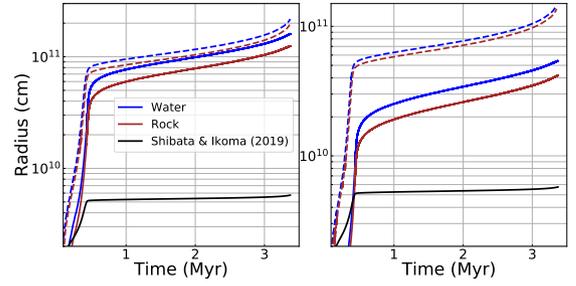}
	\caption{$R_\mathrm{capt}$ as a function of time.
	In the left and right panel plot the planetesimal's size is assumed to be 1 km and 100 km, respectively. 
	Different colours represent different planetesimal compositions where the blue and brown lines correspond to planetesimals made of water and rock, respectively. 
	The solid lines correspond to the result of \protect\cite{Inaba03} while the dashed ones correspond to Eq.~\ref{finalapproximation}.
	}
	\label{Fig6-composition}
\end{figure}

The sensitivity of $R_\mathrm{capt}$ on the planetesimal's composition is shown in Figure \ref{Fig6-composition}. 
The left and right panels show the results when assuming planetesimal sizes of 1 km and 100 km, respectively. 
We compare the inferred $R_\mathrm{capt}$ for planetesimals composed of pure-rock and pure-water. 
Rock is represented by SiO$_2$, with a material density of 2.7 g cm$^{-3}$. 
The solid lines indicate the result of \cite{Inaba03} using the density profile obtained numerically, while the dashed ones represent our Eq.~\ref{finalapproximation}.
As expected, planetesimals of lower (higher) density have a larger (smaller) capture radius.  
Different assumed planetesimal compositions would slightly affect our approximation via the parameter $\rho_\star$ introduced in Equation \ref{rhostar}, and indeed as shown in Figure \ref{Fig6-composition}, $R_\mathrm{capt}$ depends on the assumed planetesimal's composition. 
There is a very good agreement between the full numerical results and our approximation for the different assumed planetesimal's composition.
It can be seen that the difference in $R_\mathrm{capt}$ between different assumed  planetesimal's composition is significantly smaller in comparison to the typical approximations assuming the same planetesimal composition used by \cite{Shibata1}.

\section{Detached phase}
\label{appendix2}
\begin{figure}
	\centering
	\includegraphics[width=0.8\linewidth]{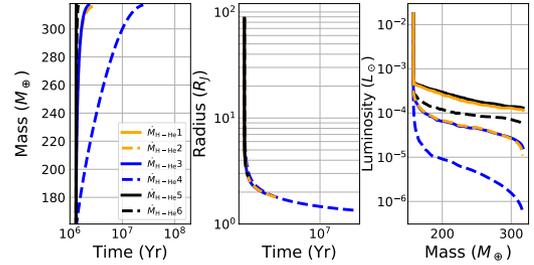}
	\caption{\textbf{Left:} Planetary mass as a function of time for  different runaway gas accretion rates. \textbf{Middle:} Radius of the planet (in Jupiter's radii units) as a function of time. \textbf{Right:} Planetary luminosity as a function of time. The different runaway gas accretion rates are defined in \protect\cite{Valletta2020} }
	\label{Fig7-mdothhe}
\end{figure}

In this section we show that the planetary radius does not depends on the assumed $\dot{M}_{xy}$. 
Figure \ref{Fig7-mdothhe} shows the radius for different assumed runaway gas accretion rates. 
The definition of the different rates can be found in \cite{Valletta2020}.
The left panel shows the total planetary mass vs.~time. There is a large difference between the different gas accretion rates, and in addition,  different gas accretion rates lead to different planetary luminosity, as shown in the right panel.
The middle panel plot shows the radius as a function of time:  all the different gas accretion rates results in a similar radius. This is similar to what found in Figure 1 of \cite{Ginzburg2019}. 

\end{document}